# Quantum walks of correlated photons in non-Hermitian photonic lattices

Mingyuan Gao[1†], Chong Sheng[1,4†*], Yule Zhao[1†], Runqiu He[1], Liangliang Lu[2], Wei Chen[1], Kun Ding[3*], Shining Zhu[1] and Hui Liu[1*]

[1]National Laboratory of Solid State Microstructures and School of Physics, Collaborative Innovation Center of Advanced Microstructures, Nanjing University, Nanjing, Jiangsu 210093, China.
[2]Key Laboratory of Optoelectronic Technology of Jiangsu Province, School of Physical Science and Technology, Nanjing Normal University, Nanjing, 210023, China.
[3]Department of Physics, State Key Laboratory of Surface Physics, and Key Laboratory of Micro and Nano Photonic Structures (Ministry of Education), Fudan University, Shanghai 200438, China.
[4]Key Laboratory of Quantum Materials and Devices (Southeast University), Ministry of Education, Nanjing, Jiangsu 211189, China.
*E-mail: csheng@nju.edu.cn
*E-mail: kunding@fudan.edu.cn
*E-mail: liuhui@nju.edu.cn
†Equally contributed to this work.

**Abstract:** Entanglement entropy characterizes the correlation of multi-particles and unveils the crucial features of open quantum systems. However, the experimental realization of exploring entanglement in non-Hermitian systems remains a challenge. In parallel, quantum walks have offered the possibility of studying the underlying mechanisms of non-Hermitian physics, which includes exceptional points, the non-Hermitian skin effect, and non-Bloch phase transitions. Unfortunately, these studies have only involved and prevailingly focused on the behavior of a single particle. Here, we propose and experimentally realize quantum walks of two indistinguishable photons in engineered non-Hermitian photonic lattices. We have successfully observed the unidirectional behavior of quantum walks in the bulk far from the edges induced by the skin effect. Moreover, we experimentally reveal the suppression of entanglement that is caused by the skin effect in non-Hermitian systems. Our study may facilitate a deep understanding of entanglement in open quantum many-body systems that are far from thermal equilibrium.

# I. INTRODUCTION

Open quantum systems are ubiquitous in nature and possess unique and complex features unknown to their closed counterparts. This has led to the development of the non-Hermitian theory [1-5]. The non-Hermitian theory has permeated various physical systems, including photonics [6-10], acoustics [11-14], cold atoms [15-17], and topolelectrical circuits [18-20], resulting in significant consequences and promising applications, such as unidirectional invisibility [21,22], high-performance lasers [23-25], enhanced sensing [26,27], and topological energy transfers [28,29], *etc*. Among these, photonic quantum walks [30] have emerged as a competent platform to study the underlying mechanism of non-Hermitian physics. The notable examples include the non-Hermitian skin effect (NHSE) [31], non-Bloch topological invariants [32], non-Bloch parity-time symmetry and phase transitions [33-35]. Despite these remarkable advances, it is essential to note that all of these studies have been carried out through the quantum dynamics of a single-photon wave packet, which can be explained classically. Notably, when the evolution of quantum walks involves more than one particle [36-39], their dynamics exhibit a hallmark feature of multiparticle interference and lack a classical analog. A natural question then arises: What are the effects of non-Hermitian physics on the quantum walks of multiple particles? Unfortunately, few experiments have been conducted to investigate the quantum walks of many bodies in non-Hermitian systems.

Furthermore, non-Hermitian theories provide a profound understanding of many-body behaviors in open quantum systems, including the dynamics of quantum correlation and entanglement among many-body particles [40-42]. Although there have been theoretical studies on the suppression of entanglement induced by the intrinsic NHSE in open condensed matter systems [42], the experimental realization of exploring this entanglement dynamic in such systems remains challenging. On the other hand, passive linear photonics systems, unlike condensed matter systems with untamed and dazzling electron interactions, offer unique advantages in the measurement of multiphoton behavior. They have been the controllable platforms for exploring non-

Hermitian behaviors [43,44]. Nonetheless, there have been no reports on experimentally studying the quantum dynamics of correlated photons in an engineered non-Hermitian system.

In this work, we propose the use of silicon-on-insulator (SOI) technology to construct a non-Hermitian system comprising silicon waveguide arrays that can manipulate non-Bloch behavior. To exhibit the non-Hermitian feature of this engineered photonic lattice, we first experimentally realize quantum walks of single photons and observe the unique non-Hermitian evolution behavior of the wave dynamics of single photons in the bulk far from the edge. Crucially, we next study the quantum walks of two indistinguishable photons in the bulk and reveal the impact of non-Hermitian properties on the dynamics of quantum correlations. We finally explore the suppression of entanglement entropy of correlated photons in non-Hermitian systems and demonstrate that the nature of non-Hermicity plays a central role in entanglement evolution, which has been verified both experimentally and theoretically.

## II. THEORY

**A. The engineered non-Hermitian photonic lattices**

To obtain a non-Hermitian photonic lattice, we judiciously design a dissipative auxiliary waveguide that links two straight waveguides, allowing for asymmetric coupling just as illustrated in Fig. 1(a). Considering that the coupling coefficient between waveguides exponentially decreases as the distance increases, these straight waveguides have negligible couplings when there is sufficient spacing between them within the limited propagation distance. Nonetheless, we can achieve equivalent coupling between these originally uncoupled straight waveguides with the aid of the auxiliary waveguide. The shape of the auxiliary waveguide can be described as $R \sin(\Omega z + \varphi)$, where $R$, $\Omega$, and $\varphi$ represent the amplitude, frequency, and geometric phase of the engineered auxiliary waveguide, respectively. The variable $z$ denotes the propagation distance. By introducing an additional loss through cutting off at the end of each auxiliary waveguide, we can control the equivalent asymmetric nearest-neighbor coupling between the straight waveguides, as depicted in the right panel of Fig. 1(a). For the case of light hopping from the left straight waveguide to the right one, and vice versa, the auxiliary waveguides in the middle capture different energy during one coupling period [45-49]. In short, we utilize the additional loss in the auxiliary waveguide to create a non-Hermitian photonic lattice with intentionally designed asymmetric coupling.

In order to determine the asymmetricity extent in the non-Hermitian photonic lattice, we employ the Lyapunov exponent [34]. And the Lyapunov exponent is given by $\lambda = \lim\limits_{z \to \infty} \frac{\log|\psi_m(z)|}{z}$, where $m$ is the site of the initial excitation waveguide. We thus numerically calculate the Lyapunov exponent for different geometric phases and give the results in Fig. 1(c). From these results, it is evident that the Lyapunov exponents are a monotonic function of the geometric phase $\varphi$ in a range from 0 to $\pi/2$. Note that the Lyapunov exponent describes the asymptotic growth rate of light intensity at the excitation position [34], which indicates the shifting behavior of the wave packet in the lattices as it propagates. To visualize the behavior of the wave packet in the lattice, we show the dynamics of the wave packet for the case with $\varphi = 0$ ($\varphi = \pi/2$) in the top panel of Fig. 1(b) [Fig. 1(d)], which illustrates that the wave packet exhibits unidirectional (unitary) diffusion in the photonic lattice with a non-zero (zero) Lyapunov exponent.

**B. The master equation for correlated photons in open systems**

As one of the most intriguing features of quantum mechanics, entanglement describes nonlocal correlations between quantum objects and lies at the heart of quantum information science. To investigate the entanglement of correlated photons in the open system, we use the density matrix, which requires the master equation to describe the evolution of multiparticle [50,51]. Considering our system, we treat the loss as the interaction with the outer environment and then introduce the Hamiltonian of the total system, which includes both the environment and the interaction. After some algebraic processes in Appendix A, the density matrix ($\rho$) governed by the Lindblad master equation is given as:

$$\dot{\rho} = -i[H',\rho] + \sum_n a_n \rho(t) a_n^\dagger - \frac{a_n^\dagger a_n}{2}\rho(t) - \rho(t)\frac{a_n^\dagger a_n}{2}, \quad (1)$$

where $H' = -\Delta\omega \sum_n a_n^\dagger a_n$, $a_n$ ($a_n^\dagger$) stands for the annihilation (creation) operator of straight waveguides. The solution of the Lindblad master equation can be obtained by mapping the operator in Fock space to the matrix in an extended linear space [50] (see details in Appendix A).

For the photonic lattice system under consideration, we adopt second-order Rényi entropy to depict the quantum entanglement of correlated photons in such non-Hermitian systems [52-54] because it extracts information about the entanglement attributes from the density matrix [51]. The Rényi entropy is defined as $S_2 = -log Tr\rho_A^2$, where $\rho_A$ is the reduced density matrix after tracing out one particle [51]. Given the initial state $\rho(0) = a_5^\dagger a_6^\dagger |0\rangle\langle 0| a_5 a_6$, the evolution of Rényi entropy of two indistinguishable photons by solving Eq. (1) is shown in Fig. 1(e), and it grows from zero during the propagation, regardless of whether the coupling is symmetric or asymmetric. The exciting behavior shows explicitly after sufficiently long periods of propagation ($z > 10T$): the entropy in the asymmetric photonic lattice ($\lambda \neq 0$) becomes lower than that in the symmetric one ($\lambda = 0$), and such a non-Hermitian system with a larger amplitude of Lyapunov exponent more efficiently suppresses the entropy compared to the symmetric system. Those behaviors suggest that the entanglement suppression of correlated photons has a close relation to the asymmetricity of the system, while such asymmetry also leads to the presence of NHSE.

## C. Entanglement entropy suppressed by NHSE

To reveal the role of NHSE in the entanglement entropy evolution [Fig. 1(e)], we employ the concept of the effective Hamiltonian and the generalized Brillouin zone (GBZ) [2]. We firstly utilize the transmission matrix $U$ of the single photon [2], and the effective Hamiltonian is then defined as $U(T,0) = e^{-i\int_0^T H(t)dt} = e^{-iH_{eff}T}$ (see discussions in Appendix B). The Schrodinger equation governed by the effective Hamiltonian originates from the semiclassical approximation of the master equation.

The results of entropy evolution calculated from the effective Hamiltonian [Fig. 1(e)] indicate that such an effective Hamiltonian method is an appropriate approximation compared to the master equation method as far as the quantum walks of correlated photons are concerned.

The GBZ framework is then an approach to analyze the impact of NHSE in the Rényi entropy, and the GBZ shape is determined by the effective Hamiltonian obtained above. Note that a system with the NHSE will render GBZ as a nonunit circle on the complex $\beta$ plane [Fig. 1(f)], and the NHSE leads to unidirectional propagation in bulk [34]. Then, we compare the energy spectrum of these effective Hamiltonians for both periodic boundary conditions (PBCs) and open boundary conditions (OBCs) as depicted in the bottom panels of Figs. 1(b) and 1(d). For all geometric phases except $\varphi = \pi/2$, the energy spectrum for PBCs forms closed loops and encloses a nonvanishing area. However, for the case with $\varphi = \pi/2$, the energy spectrum for OBCs becomes a repeated straight line. This suggests that the NHSE disappears as the phase $\varphi$ approaches $\pi/2$. Additionally, the shape of the GBZ deviates from a unit circle ($|\beta| < 1$) for all phases except $\varphi = \pi/2$. From these observations, we conclude that the magnitude of the asymmetric hopping and the resulting unidirectional diffusion in the non-Hermitian photonic lattice can be controlled by tuning the geometric phase $\varphi$.

To unveil the critical role of the NHSE in entanglement entropy suppression, we have exploited the GBZ to transform the effective non-Hermitian Hamiltonian into the one without skin effect. The GBZ of the original lattice model presented in Fig. 1(f) shows the evenly shrunken (or expanded) circle compared to the Bloch-Hamiltonian, it naturally leads to the similarity transform:

$$\overline{H} = SH_{eff}S^{-1}, S = diag\{e^{-1g}, e^{-2g}, e^{-3g}, \dots, e^{-ng}\}, (2)$$

where the skin depth $g$ is related to the radius of the GBZ circle as $e^g = |\beta_{1(2)}|$.

After performing the transformation on the OBC Hamiltonian, we can rotate the spectrum by multiplying $i$ on the transformed Hamiltonian to make it Hermitian:

$$\widetilde{H} = i\overline{H}, (3)$$

$$\widetilde{H}|\overline{\psi_n}\rangle = iE_n|\overline{\psi_n}\rangle, (4)$$

The similarity transformation contains the effect of NHSE, and the $\widetilde{H}$ (or $\overline{H}$) can interpret the lossy and symmetric hopping lattice model.

We compare the evolution of two indistinguishable photons in both the original and the transformed Hamiltonian system [see dashed lines in Fig. 1(e)]. The entropy evolution of the two systems shows distinct behavior with or without the NHSE. The entropy is no longer suppressed in the transformed system. This clearly indicates that the skin effect plays an essential role in manipulating the entropy of multiphotons walking in non-Hermitian systems: in the asymmetric lattices, such suppression behavior of Rényi entropy originates from that the correlated photons bunch toward the edge of the lattice due to the skin effect; in contrast, for the symmetric lattices the two photons just dissipate over the real space, and the Rényi entropy will increase and eventually reach a constant value. Importantly and fortunately, the Rényi entropy as the entanglement indicator can be observed through quantum walks of correlated photons, and our theory can then be validated by the experimental results below.

## II. EXPERIMENT

**A. Quantum walks of single photon and skin effect**

In experiments, we use commercial SOI technology to fabricate the non-Hermitian photonic lattice. To study the evolution of photons within the lattice, we use multiple straight waveguides as sources to inject photons into the lattice and select all the straight waveguides as output sources for detecting the photons after passing through the lattice. Considering that we leverage the loss to achieve asymmetric coupling, the evolution period in experiments cannot reach a large value. Thus, we fabricate a series of evolution periods with $z = 3T, 4T, 5T, 6T$ for various types of non-Hermitian lattices, and the number of the straight waveguide is $N = 9$. Additionally, different types of quantum light sources are prepared using spontaneous parametric down-conversion by pumping a Type II periodically poled lithium-niobate (PPLN) waveguide (see detailed methods in Appendix C). For all the data of coincidence detections in experiments, accidental coincidence counts are subtracted.

First, we measure the quantum walks of single photons in the non-Hermitian photonic lattice. Figure 2 shows the quantum walks of single photons for injecting photons into the same waveguide site ($n_0 = 6$) in the non-Hermitian photonic lattice under different Lyapunov exponents and evolution periods. Moreover, we compare the experimental results with theoretical results using the similarity measure given by $Q = \left(\sum_n \sqrt{p_n^{(exp)} p_n^{(thr)}}\right)^2 \bigg/ \left(\sum_n p_n^{(exp)} \sum_n p_n^{(exp)}\right)$, where $p_n = \langle c_n^+ c_n \rangle = |U_{n,n_0}|^2$, and $U_{n,n_0}$ is the transmission matrix indicating the amplitude for the transition of a single photon from the site $n_0$ to site $n$. And we find that the least similarity is still up to $85.3\% \pm 1.9\%$. From the measurement of the probability distribution, the unidirectional movement of the wave pack clearly proves that such photonic lattices with a nonzero Lyapunov exponent possess the NHSE. As we expected from the simulation above, the unidirectional shift distance depends on the Lyapunov exponent. There is no shift of the wave packet of single photons in the case of zero Lyapunov exponent [Fig. 2(m)]. In contrast, when the magnitude of the Lyapunov exponent increases, there is a more significant shift of the wave packet during propagation [Fig. 2(n) and Fig. 2(o)].

**B. Quantum walks of correlated photons and entanglement suppression**

Next, we study the quantum interferences of correlated photons by inspecting the two-photon correlation distribution. In theory, we solve the Eq. (1) numerically in the photonic lattice with different Lyapunov exponents and extract the correlation matrix $\Gamma_{nm}(t)$. The definition is $\Gamma_{nm}(t) = \langle n, m | \rho(t) | n, m \rangle$, where $n$ and $m$ indicate the site position of the photons. We simulate the two-photon quantum walks with two indistinguishable photons excitation at the sites $n_0 = 5$ and $m_0 = 6$, meaning that the initial state is $\rho(0) = a_5^\dagger a_6^\dagger |0\rangle\langle 0| a_5 a_6$. In the experiment, we inject the photonic waveguide array with two indistinguishable photons at corresponding positions and measure the correlation matrix $\Gamma_{nm}(t)$ under various Lyapunov exponents. We achieve the measurement of the correlation probability distribution for evolutionary periods in $3 \sim 6T$.

The comparison between theoretical and experimental results is shown in Figs. 3(a)–(l). We find that the patterns of the coincident probability distribution shift towards the corner in the case of a nonzero Lyapunov exponent. Additionally, the shifting behavior of the entire pattern becomes more pronounced in cases with a larger magnitude of Lyapunov exponent and longer evolution periods. Obviously, such behavior of the coincident probability distribution arises from the unidirectional diffusion in the bulk caused by the NHSE, which is akin to that of quantum walks of single photons. Moreover, the similarity between the two matrices of the simulated and measured ones, defined by $Q = \left(\sum_{nm}\sqrt{\Gamma_{nm}^{(exp)}\Gamma_{nm}^{(thr)}}\right)^2 \Big/ \left(\sum_{nm}\Gamma_{nm}^{(exp)}\sum_{nm}\Gamma_{nm}^{(thr)}\right)$, validate that the experimental results are in good agreement with the theoretical results.

Significantly, the Rényi entropy can be retrieved from the coincident matrix [52,53] (see details in Appendix D), which can be observed directly in the experiment. Figure 3(m) shows the evolution of the Rényi entropy of theoretical and experimental results for evolutionary periods in $3\sim6T$. To clearly illustrate the entropy affected by the NHSE, the entropy in the asymmetric photonic lattice is normalized by the symmetric one as: $S_{norm} = S_2 - S_2|_{\lambda=0}$. The experimental results indicate that the entanglement entropy suppression exists even in the cases of relatively short evolutionary periods: the Rényi entropy in the asymmetric photonic lattice ($\lambda \neq 0$) is lower than that in the symmetric one ($\lambda = 0$), and the entropy in the asymmetric system is suppressed more heavily as the evolution proceeds. Although the measurement of correlation for longer evolution periods is not achievable in current experiments due to the lossy signals, the extended theoretical results [see Fig. 3(n) for the normalized ones and Fig. 1(e) for the original ones] show that the suppression of entropy emerges more profoundly as the propagation continues. These results further indicate the crucial role of NHSE on the entropy evolution when correlated photon walks among non-Hermitian lattices.

## IV. DISCUSSION

In conclusion, we have experimentally realized quantum walks of single photons and two indistinguishable photons in engineered non-Hermitian photonic lattices. We strategically utilize the dissipative auxiliary waveguides to manipulate asymmetric coupling. The unidirectional behavior of quantum walks caused by the skin effect has been observed. Moreover, we experimentally study the dynamics of quantum entanglement of correlated photons in non-Hermitian systems and observe the suppression of entanglement induced by the skin effect. Our work exploits quantum walks of correlated photons in silicon photonics as a platform for studying multiparticle non-Hermitian physics. Currently, we utilize the second-order Rényi entropy only considering diagonal elements of the density matrix to depict quantum entanglement of correlated photons. In the future, with the development of the precise phase measurement of photons, one can experimentally reconstruct the complete information of the density matrix, which can be explored to depict correlated behaviors of multiple particles using various types of entropy. Moreover, our experimental platform may be utilized to explore more challenging problems in open quantum systems. Since silicon itself possesses good nonlinear optical properties, studies covering sophisticated topics, such as nonlinearity and non-Hermiticity, have the possibility of being conducted on our experimental platform.

## ACKNOWLEDGEMENTS

This work was financially supported by National Key Research and Development Program of China (2023YFB2805700), the National Natural Science Foundation of China (Grant Nos. 12174187, 62288101, 92163216, 92150302, 2021hwyq05), the Natural Science Foundation of Jiangsu Province, China (BK20240164, BK20243009) and the Fundamental Research Fund for the Central Universities, China (Grant No. 2024300329). C.S. acknowledges additional support from the open research fund of Key Laboratory of Quantum Materials and Devices (Southeast University), Ministry of Education.

**APPENDIX A: MASTER EQUATION FORMALISM**

In this section, we discuss the general mathematical method for calculating the transmission behavior of our dissipative model. Given that we are examining an open quantum system, we employ the master equation of the density matrix [50,55], and discuss the numerical method for its solution.

In our model, light propagation can be separated into two distinct processes within a single period. For the phenomena of interest, we consider the entire waveguide array (including both straight and auxiliary waveguides) as the system, with the remaining components of the SOI system treated as the environment.

We begin by analyzing the first process, corresponding to the propagation part in Fig. 4. The light, initialized in a specific state, propagates through both straight and auxiliary waveguides until the auxiliary waveguides are truncated. During this process, the system and the environment are completely decoupled, rendering the density operator of the total system separatable. The evolution of the density operator during this process is governed solely by the Hamiltonian of the system, which is equivalent to the standard Hamiltonian formalism.

The second process, which we refer to as the dissipation part in Fig. 4, commences once the auxiliary waveguides are cut off and continues until the start of the next period. Photons within the auxiliary waveguide dissipate into the environment. Thus, the auxiliary waveguides weakly interact with the environment (indicated by the dashed waveguide in Fig. 4) while remaining decoupled from the straight waveguides. Additionally, we assume that the environment consists of a large resonant chamber containing multiple electromagnetic eigenmodes, with which the auxiliary waveguides may interact.

Specifically, the total system has been divided into the system and the environment, and can be described by the evolution equation of the density operator in the Schrodinger picture:

$$\dot{\rho}_{tot} = -i[H_{tot}, \rho_{tot}], H_{tot} = H_S + H_E + V. \text{(A1)}$$

The $H_S$, $H_E$, and $V$ are the Hamiltonian of the system, the environment, and the interaction between them, respectively.

For simplicity, the assumption of Hamiltonian of the system is depicted as:

$$H_S = \beta_0 \sum_j a_j^\dagger a_j, \quad (A2)$$

where the subscript $j$ denotes only the index of auxiliary waveguides. The Hamiltonian of the environment is depicted as:

$$H_E = \sum_k \omega_k b_k^\dagger b_k, \quad (A3)$$

where $b_k^\dagger$ ( $b_k$ ) represents the creation (annihilation) operator of different electromagnetic modes in the environment. The interaction between the system and environment is depicted as:

$$V = \sum_{k,j} (g_k b_k + g_k b_k^\dagger)(a_j + a_j^\dagger). \quad (A4)$$

Considering that there is no nonlinear term in our system, we should remove the $g_k b_k^\dagger a_j^\dagger$ and $g_k b_k a_j$ terms above. After transforming into the interaction picture, we obtain:

$$\dot{\rho}_{tot,I}(t) = -i[V_I, \rho_{tot,I}(t)], \quad (A5)$$

$$V_I(t) = \sum_{k,j} \left( g_k e^{-i(\omega_k - \beta_0)t} b_k a_j^\dagger + g_k e^{i(\omega_k - \beta_0)t} b_k^\dagger a_j \right). \quad (A6)$$

After substitution and integration twice, we get the motion equation in integration form [55]:

$$\dot{\rho}_{tot}(t) = -i[V(t), \rho_{tot}(0)] - \int_0^t dt_1 \, [V(t), [V(t_1), \rho_{tot}(t_1)]]. \quad (A7)$$

Omitting the subscription $I$ for simplicity and tracing over the environment, denoting $\rho = \mathrm{Tr}_E \rho_{tot}$, we obtain

$$\dot{\rho}(t) = -i Tr_E[V(t), \rho_{tot}(0)] - \int_0^t dt_1 \, Tr_E[V(t), [V(t_1), \rho_{tot}(t_1)]]. \quad (A8)$$

We can substitute the above term into the density matrix equation and assume that the initial state of the environment is the vacuum state. The first term of the right-hand side will be zero:

$$\text{Tr}_E[V(t), \rho_{tot}(0)] = 0. \quad (A9)$$

The second term will be simplified into

$$-\int_0^t dt_1 \sum_{j,j'} [\Gamma(t-t_1)(\rho(t_1)a_j^\dagger a_{j'} - a_{j'}\rho(t_1)a_j^\dagger) + h.c.], \quad (A10)$$

where $\Gamma(\tau) = \sum_k g_k^2 e^{i(\omega_k - \beta_0)\tau} \simeq \int dk \rho(k) g^2(k) e^{i(\omega_k - \beta_0)\tau}$. Because the environment works as a resonant chamber, which is relatively much larger than the cross-face of the waveguides, the spectrum of the mode is approximately continuous, and one can change the summation into integration. The integrand with the rapid oscillation term $e^{i(\omega_k - \beta_0)\tau}$ leads to that $\Gamma(\tau)$ sharply peaked at $\tau = 0$. The integrand in the second term should contribute mainly at $t_1 = t$. The motion equation of the density matrix will become

$$\dot{\rho} = -i\Delta\omega \left[\rho, \sum_{j,j'} a_j^\dagger a_{j'}\right] + \sum_{j,j'} 2\gamma a_j \rho a_{j'}^\dagger - \gamma\{\rho, a_j^\dagger a_{j'}\}, \quad (A11)$$

where $\gamma + i\Delta\omega = \int_0^\infty \Gamma(\tau) d\tau$. Here, we ignore the $j \neq j'$ terms in the summation which represent the inter-coupling between the auxiliary waveguides with the help of the environment reservoir.

The motion equation of the density matrix of the waveguide array becomes the form of Lindblad master equation, which describes the loss process from the effective Hamiltonian [50]:

$$\dot{\rho} = -i[H', \rho] + \sum_j D[\sqrt{2}a_j]\rho(t), H' = -\Delta\omega \sum_j a_j^\dagger a_j. \quad (A12)$$

The dissipator is defined as $D[A]\rho = A\rho A^\dagger - \frac{A^\dagger A}{2}\rho - \rho\frac{A^\dagger A}{2}$.

After getting the master equation of the dissipation process, we introduce the dissipation strength $\gamma$:

$$\dot{\rho} = -i[H', \rho] - i\sum_j (-i\gamma a_j^\dagger a_j \rho - i\gamma \rho a_j^\dagger a_j) + 2\gamma a_j \rho a_j^\dagger$$

$$= -i(\tilde{H}\rho - \rho\tilde{H}^\dagger) + 2\gamma \sum_j a_j \rho a_j^\dagger, \quad (A13)$$

where $\tilde{H} = H' - i\gamma \sum_j a_j^\dagger a_j$, and the uniform onsite potential in $H'$ can be omitted. We can find that $\widetilde{H}$ leads to the loss of the auxiliary waveguides.

Utilizing quantum jump theory, we first proved that only $m$-particle states ($m < n$) in the Hilbert space will be possessed since the initial state only has $n$ particles. Assuming the initial state as an arbitrary complex state:

$$\rho = \sum_k c_k |\psi_k\rangle\langle\psi_k|. \quad (A14)$$

The master equation is

$$\partial_t \rho = \sum_k -ic_k\big(\widetilde{H}|\psi_k\rangle\langle\psi_k| - |\psi_k\rangle\langle\psi_k|\widetilde{H}^\dagger\big) + 2\gamma \sum_{j,k} a_j c_k |\psi_k\rangle\langle\psi_k| a_j^\dagger. \quad (A15)$$

The first term describes the pure state $|\psi_k\rangle$ evolution under the non-Hermitian Hamiltonian $\widetilde{H}$; the second term shows the composition of the state after the quantum jump.

One can denote the time-evolution of the pure state under non-Hermitian Hamiltonian as $|\psi_k(t)\rangle$, and define the state after the quantum jump $|\varphi_k\rangle_j = a_j |\psi_k\rangle$.

The dynamic process governed by the Hamiltonian shows

$$|\psi_k(t + \delta t)\rangle = (1 - i\widetilde{H}\delta t)|\psi_k(t)\rangle. \quad (A16)$$

The normalization is

$$\langle\psi_k(t + \delta t)|\psi_k(t + \delta t)\rangle = \langle\psi_k(t)|\big(1 - i(\widetilde{H} - \widetilde{H}^\dagger)\delta t\big)|\psi_k(t)\rangle$$

$$= \langle\psi_k(t)|\big(1 - i(-2\gamma i \sum_j a_j^\dagger a_j)\delta t\big)|\psi_k(t)\rangle$$

$$= 1 - \sum_j \delta p_j, \quad (A17)$$

where $\delta p_j = 2\delta t \gamma \langle\varphi_k|\varphi_k\rangle_j$ describes the quantum jump possibility affected by the $j_{th}$ auxiliary waveguide. Moreover, the density operator shows

$$\rho(t + \delta t) = \sum_k c_k \Big(\big(1 - \sum_j \delta p_j\big)|\psi_k(t+\delta t)\rangle\langle\psi_k(t+\delta t)|$$

$$+ \sum_j \delta p_j |\varphi_k\rangle_j\langle\varphi_k|_j\Big). \quad (A18)$$

The first term describes the loss governed by the non-Hermitian Hamiltonian $\widetilde{H}$, which introduces lossy onsite energy on the auxiliary waveguides, with the possibility $(1 - \sum_j \delta p_j)$. The second term describes the quantum jump process of elimination photon at $j_{th}$ auxiliary waveguide with the possibility $\delta p_j$. The density operator cannot possess a state with more particles than the initial state. After a sufficiently long period of dissipation, all the state components in the density matrix will ultimately decay to the vacuum state.

For the concern we are studying, we solve the master equation considering the Fock state component only containing the state no more than two photons:

$$\rho = c_0|0\rangle\langle 0| + \sum_n c_{n,1}|1_n\rangle\langle 0| + \sum_j c_{j,2}|0\rangle\langle 1_j| + \sum_{n,j} c_{n,j,3}|1_n\rangle\langle 1_j|$$
$$+ \sum_{n,m} c_{n,m,4}|2_{n,m}\rangle\langle 0| + \sum_{j,l} c_{j,l,5}|0\rangle\langle 2_{j,l}|$$
$$+ \sum_{n,m,j} c_{n,m,j,6}|2_{n,m}\rangle\langle 1_j| + \sum_{n,j,l} c_{n,j,l,7}|1_n\rangle\langle 2_{j,l}|$$
$$+ \sum_{n,m,j,l} c_{n,m,j,l,8}|2_{n,m}\rangle\langle 2_{j,l}|. \text{(A19)}$$

For the convenience of numerical manipulation, we employ an alternative scheme that represents the density matrix $\rho$ as an extended matrix. In this scheme, the left (right) multiplication of the Hamiltonian operator $\tilde{H}$ and annihilation operator $a_i$ is treated as matrix multiplication. This allows us to represent the left (right) vector as an $N^2 + N + 1$ dimension vector, which corresponds to a state containing no more than two photons. The first $N^2$ components describe the state with two photons at site $n, m$; the next $N$ components represent the single-photon state at site $l$; and the last component corresponds to the vacuum state. An arbitrary density matrix, expanded by these basis states, can thus be represented by a matrix of dimension $(N^2 + N + 1) \times (N^2 + N + 1)$, thereby preserving all the degrees of freedom relevant to our analysis.

We can then express the density matrix of pure states as

$$\rho = \begin{bmatrix} \vdots \\ |n,m\rangle \\ \vdots \\ |l\rangle \\ \vdots \\ |0\rangle \end{bmatrix} [\cdots \quad |n',m'\rangle \quad \cdots \quad |l'\rangle \quad \cdots \quad |0\rangle]. \text{(A20)}$$

Followingly, the density matrix of an arbitrary mixed state is depicted as

$$\rho = \begin{bmatrix} [\rho]_{nm,n'm'} & [\rho]_{nm,l'} & [\rho]_{nm,0} \\ [\rho]_{l,n'm'} & [\rho]_{l,l'} & [\rho]_{l,0} \\ [\rho]_{0,n'm'} & [\rho]_{0,l'} & \rho_{00} \end{bmatrix}. \text{(A21)}$$

The master equation in a single period is separated into two processes:

Propagation part: $\dot{\rho} = -i(H(t)\rho - \rho H(t))$, (A22)

Dissipation part: $\dot{\rho} = -i(\widetilde{H}\rho - \rho\widetilde{H}^{\dagger}) + 2\gamma \sum_j a_j \rho a_j^{\dagger}$, (A23)

The Hamiltonian in the new scheme becomes the extended matrix shown below:

$$H \Rightarrow \begin{bmatrix} H \otimes I + I \otimes H & 0 & 0 \\ 0 & H & 0 \\ 0 & 0 & 0 \end{bmatrix},$$

and the annihilation and creation operators in the new scheme will be the matrix as

$$a_j \Rightarrow \begin{bmatrix} 0 & 0 & 0 \\ [T_{l,nm}] & 0 & 0 \\ 0 & [R_l] & 0 \end{bmatrix},$$

$$a_j^{\dagger} \Rightarrow \begin{bmatrix} 0 & 0 & 0 \\ [T_{l,nm}] & 0 & 0 \\ 0 & [R_l] & 0 \end{bmatrix}^T,$$

where $T_{l,nm} = \frac{1}{1+\delta_{nm}}(\delta_{jn}\delta_{lm} + \delta_{jm}\delta_{ln})$, $R_l = \delta_{jl}$.

Starting with an initial state of $\rho(0)$, the time-evolved state $\rho(t)$ can be determined for each period $t = T, 2T, 3T, \ldots$ by numerically solving the corresponding ordinary differential equations in the extended linear space. From $\rho(t)$, the correlation matrix and the second-order Rényi entropy can be calculated, reflecting the behavior of correlated photons as they propagate through the dissipative waveguide lattice, as illustrated in Fig. 1(b).

**APPENDIX B: EFFECTIVE NON-HERMITIAN HAMILTONIAN**

In this section, we discuss the numerical tool used to analyze single photon transmission in a system exhibiting NHSE, which involves the Schrodinger equation of single photons in photonic waveguides. We employ the tight-binding model of a photonic lattice, which leads to the coupled-mode equations in waveguides [46-49]:

$$i\frac{dc_n}{dz} = -\kappa_{n,n+1}c_{n+1} - \kappa_{n,n-1}c_{n-1} - \beta_n c_n, \text{(B1)}$$

where $c_n$ is the (complex) amplitude in $n$-th waveguide, and $\kappa_{j,i}$ is the effective mode coupling coefficient of the directional coupler from $i$-th waveguide to $j$-th waveguide, while $\beta_n$ acts as on-site energy.

We first simulate the propagating mode (TM$_0$) in silicon waveguides using the COMSOL software. This allows us to determine the relationship between effective hopping and waveguides distance. The results indicate that the distance between two straight waveguides and the effective coupling follow an exponential relation [48]:

$$\kappa = Ae^{-bx}, A = 13.99\mu m^{-1}, b = 8.26\mu m^{-1}$$

Note that the coupling coefficient is reciprocal between straight waveguides. To get the best nonreciprocal phenomenon while keeping the robustness of the on-chip design for SOI technology, we choose the parameters of the model as $a = 0.9\mu m, R = 0.21\mu m, T = 40\mu m$.

Since the period length is long enough, we can reasonably assume that the difference in propagation direction between the straight and auxiliary waveguides is negligible. As a result, the exponential relation of the hopping persists along propagation. Provided the coupling mode equation serves as an analog to the Schrödinger equation, the Hamiltonian of the system over a single period becomes time-dependent.

$$i d \begin{pmatrix} \vdots \\ c_n \\ \vdots \end{pmatrix}/dz = H(z) \begin{pmatrix} \vdots \\ c_n \\ \vdots \end{pmatrix}, H_{nm}(z) = \begin{cases} \kappa(z), & n = m \pm 1 \\ \beta_0, & n = m \\ 0, & otherwise \end{cases} . (B2)$$

Considering the energy loss caused by the periodic cut-off in the auxiliary waveguide, all the energy at auxiliary waveguides will be lost at the ends of every period when performing a simulation. Mathematically, we need to project the transmission matrix onto the subspace containing only straight waveguides. The transmission matrix (Green function) satisfying [45]:

$$\begin{pmatrix} \vdots \\ c_n(T) \\ \vdots \end{pmatrix}_{n \in S} = U(T) \begin{pmatrix} \vdots \\ c_n(0) \\ \vdots \end{pmatrix}_{n \in S}, (B3)$$

where S denotes the indices of the subspace of straight waveguides. This subsystem, comprising the straight waveguides, is clearly non-Hermitian due to energy loss, as discussed in Appendix A. We can define the effective Hamiltonian for this non-Hermitian subsystem using the transmission matrix:

$$U(T) = e^{-iTH_{eff}}. (B4)$$

As a critical measure of NHSE, the one-dimensional GBZ [Fig. 1(f)] can be calculated from the effective Hamiltonian using a standard method [2]. Besides, the PBC and OBC spectrum can be obtained from the system with finite site numbers (N=30) as shown in Figs. 1(c) and 1(e).

Moreover, we find that the effective Hamiltonian under PBC not only exhibits nonreciprocal nearest-neighbor coupling terms but also reveals the presence of higher-order long-distance coupling terms (see details in Table I), which are not negligible.

| Coupling order | -4 | -3 | -2 | -1 | 0 | 1 | 2 | 3 | 4 | 5 |
|---|---|---|---|---|---|---|---|---|---|---|
| $Im(K)(\mu m^{-1})$ $\times 10^{-4}$ | $-3.55$ | $-9.70$ | $-29.8$ | $-123$ | $-113$ | $-12.2$ | $-0.240$ | $-0.239$ | $-0.565$ | $-1.38$ |

TABLE I. All orders of hopping in PBC effective Hamiltonian with the sites number $N = 10$ and geometric phase $\varphi = 0$. The real part is always zero for all the hopping terms.

For two-photon case, the correlation function can be defined using a reasonable approximation method with the help of the single-photon transmission matrix:

$$\Gamma_{jl}^{(n,m)} = \langle \Psi | c_j^\dagger c_l^\dagger c_j c_l | \Psi \rangle = \frac{1}{1 + |\delta_{jl}|} |U_{jn}U_{lm} + U_{jm}U_{ln}|^2, \text{(B5)}$$

where $n$ and $m$ represent the inject positions of the two indistinguishable photons. We find that the solution derived from the single-photon transmission matrix aligns closely with the results obtained using the master equation method, as demonstrated in Fig. 1(b).

## APPENDIX C: MATERIALS AND EXPERIMENTAL METHODS

### 1. Sample fabrication

In the experiments, a structured photonic lattice is fabricated by etching the device layer of an SOI wafer, with confinement provided by the buried oxide underneath and a capping oxide above. The thickness of the silicon device layer is 220 nm, while the buried oxide underneath and the capping oxide above are both 2-μm-thick silica. The waveguides are designed to be single-mode, having a width of 450 nm. The structures are defined by electron beam lithography and dry etching.

### 2. Quantum measurement

The silicon lattice contains nine straight waveguides and eight auxiliary waveguides. Labeling all the straight waveguides with index 1 to 9, we keep 3 to 7 sites on the one side as the input and all sites on the other side as the output. [Figs. 5 (c) and (d)]. For quantum walks of single photons, we choose one of photon pairs as a heralded photon injected into the designed silicon lattices for evolution, while another photon acts as a trigger signal. Meanwhile, for the quantum walk of correlated photons, two photons are simultaneously injected into designed silicon lattices for evolution [Fig. 5 (a)]. Both photons are filtered to suppress residual noise with off-chip filters and finally directed into and detected by superconducting nanowire single-photon detectors (SNSPDs). Fiber polarization controllers are used to optimize the polarization of photons for maximum detection efficiency in SNSPDs. Coincidence measurements are performed using the time-correlated single photon counting module (Picoquant PicoHarp 300).

**3. Quantum light source and measurement of HOM dip**

We generate the single-photon pair at the wavelength of 1550.92 nm via spontaneous parametric down-conversion by pumping a type-II PPLN waveguide from a continuous wave fixed at 775.46 nm. The length of the PPLN waveguide is 2 cm. The generated photon pair is separated into two components, horizontal and vertical polarization, after passing through a long-pass filter and a polarized beam splitter. Moreover, after converting the polarization of these types of single photons from the vertical state to the horizontal state, we find that the deterministically separated identical photon-pair has a very high visibility of the quantum interferences, characterized by a Hong-Ou-Mandel (HOM) dip with 97.32% ± 0.17% visibility.

**APPENDIX D: DEFINITION OF RÉNYI ENTANGLEMENT ENTROPY**

In this section, we discuss the derivation of Rényi entanglement entropy [54] from the correlation distribution measured previously. The Rényi entropy of order $n$, which depict the entanglement between two photons, is defined as:

$$E_n = \frac{1}{1-n} \log Tr\rho_A^n, \quad (D1)$$

where $\rho_A$ is the reduced density matrix, obtained by tracing out the sub-space corresponding to photon B from the two-photon density matrix: $\rho_A = Tr_B |\psi\rangle\langle\psi|$. The second-order Rényi entropy can be calculated from the correlation probability distribution, following the approach outlined in several papers [52,53].

Consider the relation of biphoton states in Fork representation:

$$|\psi\rangle = \sum_i \beta_{ii} |2_i\rangle + \sum_{i \neq j} \beta_{ij} |1_i 1_j\rangle \quad (D2)$$

For convenience, here we ignored the superscript denoting the fixed input ports $m, n$. Then, taking the trace over one of the two photons:

$$\rho_A = Tr_B \rho$$

$$= \sum_i |\beta_{ii}|^2 |i\rangle\langle i| + \sum_{i\neq j}(\beta_{ij}\beta_{jj}^*|i\rangle\langle j| + h.c.) + \sum_{i\neq j, i'\neq j} \beta_{ij}\beta_{i'j}^*|i\rangle\langle i'|, \quad (D3)$$

second-order Rényi entanglement entropy is defined as $E_2 = -\log Tr\rho_A^2$. Substitute the term with Fork representation, one can get

$$Tr\rho_A = \sum_i |\beta_{ii}|^2 + \sum_i f_i, \quad (D4)$$

$$f_i = \sum_{j=1}^{i-1} |\beta_{ji}|^2, \quad (D5)$$

$$Tr\rho_A^2 = \sum_{i=1}^N |\beta_{ii}|^4 + 2\sum_{i=1}^N f_i|\beta_{ii}|^2 + \sum_{i=1}^N f_i^2 + \sum_{i,j=1, i\neq j}^N [\rho_A]_{ij}[\rho_A]_{ji}, \quad (D6)$$

Given that in the quantum optical experiment only the intensity of each correlation term ($\beta_{ij}$) can be measured, we adopt a reasonable approximation by neglecting the summation terms where $i \neq j$:

$$Tr\rho_A^2 = \sum_{i=1}^N |\beta_{ii}|^4 + 2\sum_{i=1}^N f_i|\beta_{ii}|^2 + \sum_{i=1}^N f_i^2. \quad (D7)$$

The diagonal terms in the partial-traced density matrix are preserved, representing the components of the complex state within the partial-traced density matrix. From the probability distribution of correlation, one can calculate the reduced second-order Rényi entropy with the formula provided above.

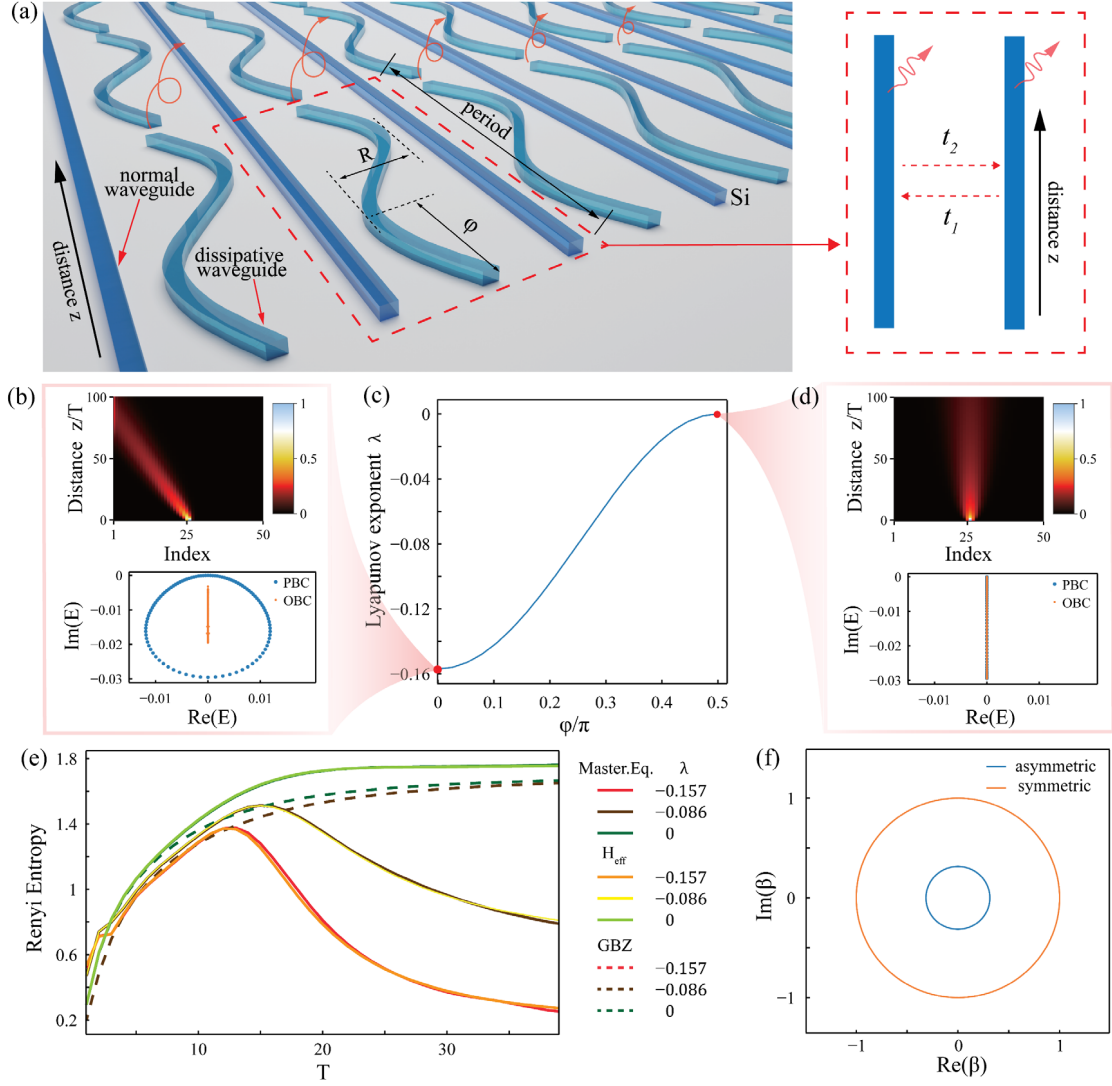

Fig. 1. The structured one-dimensional non-Hermitian lattice and Rényi entanglement entropy. (a) The designed system consists of an array of identical straight waveguides, with the auxiliary waveguide in the middle controlling their coupling. $R$ is the Floquet radius, $\varphi$ is the geometric phase, $T$ is the period length, and $z$-axis indicates the propagation direction of waveguides. On the right is the schematic of effective coupling and dissipations. (b, d) The asymmetric case with $\varphi = 0$ and the symmetric case with $\varphi = \pi/2$ respectively. (c) The relevance of the Lyapunov exponent to the geometric phase $\varphi$. (e) Rényi entropy calculated from the effective Hamiltonian and the master equation (solid lines). The corresponding results after the similarity transformation with GBZ are shown by the dashed lines. (f) The GBZ derived from the effective Hamiltonian. All the parameters used in (b) to (f) are $R = 0.21\,\mu\mathrm{m}$, $a = 0.9\,\mu\mathrm{m}$, and $T = 40\,\mu\mathrm{m}$.

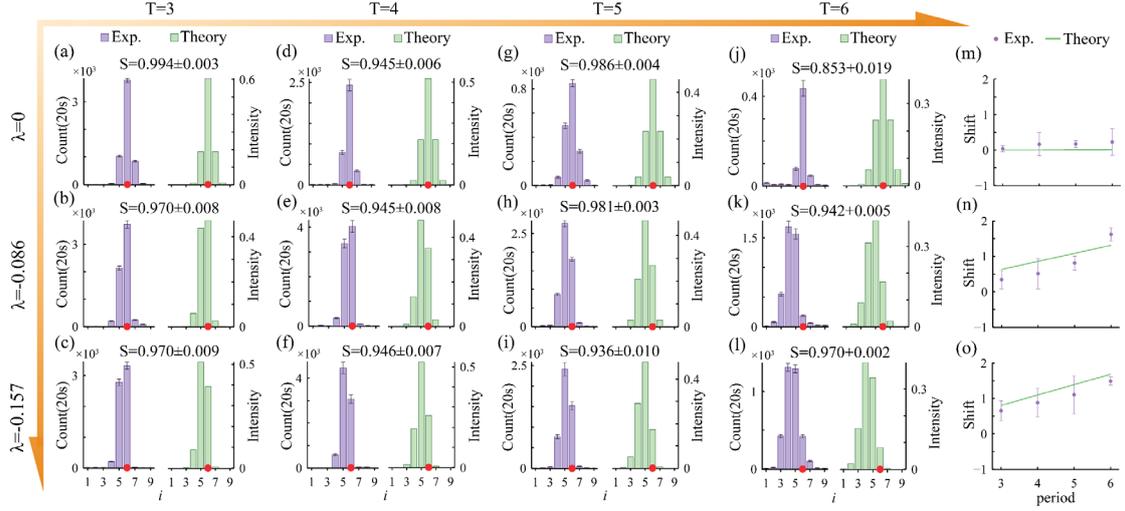

Fig. 2. Quantum walks of single photons. (a-l) The comparison of the probability of single photons between experimental (purple) and numerical (green) results under different evolution periods and Lyapunov exponents. The red dot indicates where single photons are injected. (m, n, o) The shift of the wave pack center of single photons. In experiments, the lattice system has total sites $N = 9$.

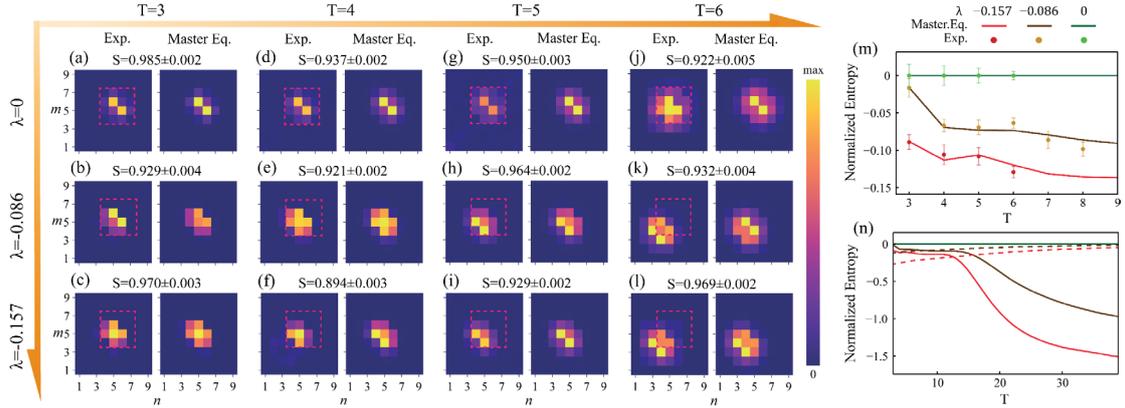

Fig. 3. The evolution of correlation distribution and Rényi entanglement entropy. (a, d, g, j) In the symmetric cases, the possibility distribution diffuses from the center. (b, e, h, k) In contrast, the whole distribution moves unidirectionally along the diagonal line due to asymmetric coupling with the Lyapunov exponent as $\lambda = -0.086$. (c, f, i, l) The evolution of the whole distribution with a larger Lyapunov exponent as $\lambda = -0.157$. The red dashed square serves as a reference background for these comparisons under various evolution periods and Lyapunov exponents. (m) The evolution of the normalized Rényi entropy under various Lyapunov exponents, including theoretical results (solid lines) calculated from the master equation and experimental results (dots). (n) The normalized entropy with an extended evolutionary period up to $40T$, and the dashed lines represent corresponding results under similarity transformation.

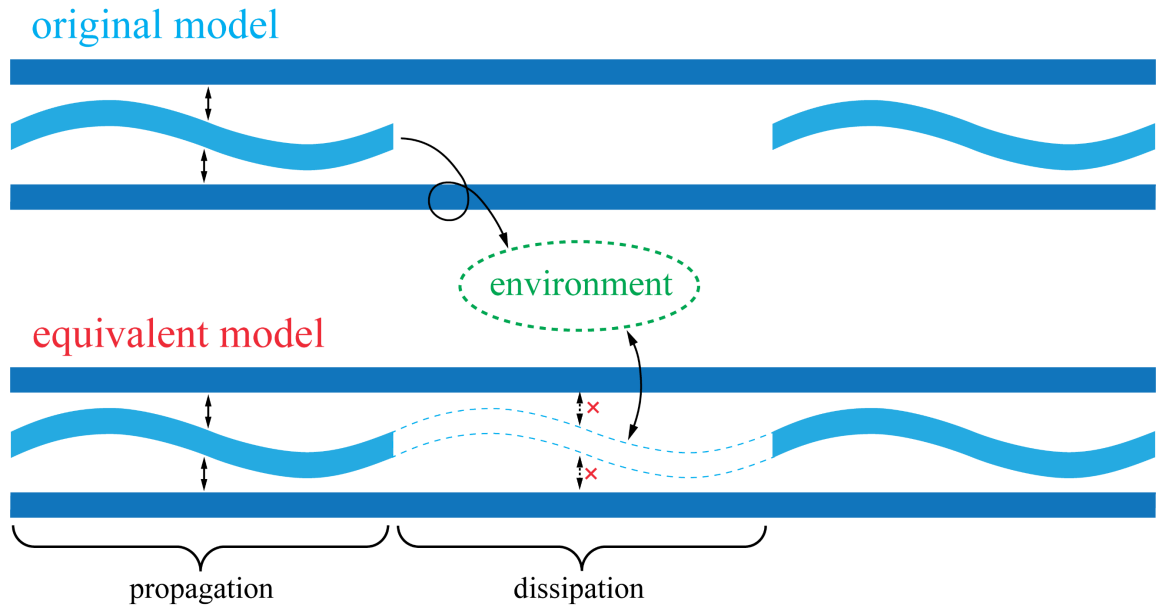

Fig. 4. Schematic for the theoretical model.

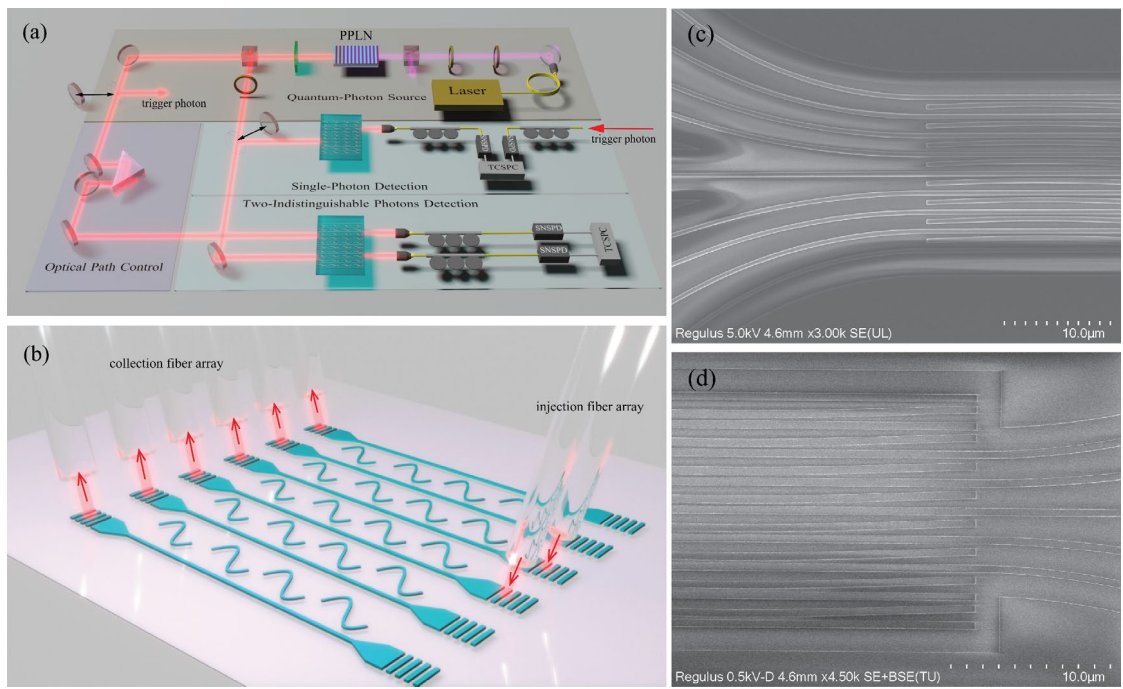

Fig. 5. (a) Schematic of the measurement of quantum walks of single photons and two indistinguishable photons. (b) Schematic of injection and collection of photons. (c, d) SEM picture of the silicon lattice. Only the straight waveguides are extended out from the lattice. (c) shows the input side, and five sites of straight waveguides are extended for photon injection. (d) shows the output side, and all nine sites of straight waveguides are extended for photon collection.